\documentclass[]{aa}
\usepackage{graphicx}
\usepackage{color}
\usepackage{txfonts}

\begin{document}

\title{Torsional Alfv\'en waves in solar partially ionized plasma: effects of neutral helium and stratification}

\author{Zaqarashvili, T.V.\inst{1,3}, Khodachenko, M.L. \inst{1} and Soler, R. \inst{2}
}

 \institute{Space Research Institute, Austrian Academy of Sciences, Schmiedlstrasse 6, 8042 Graz, Austria\\
             \email{[teimuraz.zaqarashvili;maxim.khodachenko]@oeaw.ac.at}
                               \and
           Departament de F\'{\i}sica, Universitat de les Illes Balears, E-07122 Palma de
Mallorca, Spain \\
            \email{roberto.soler@uib.es}
                                          \and
            Abastumani Astrophysical Observatory at Ilia State University, University St. 2, Tbilisi, Georgia\\
}

\date{Received / Accepted }

\abstract{Ion-neutral collisions may lead to the damping of Alfv\'en waves in chromospheric and prominence plasmas. Neutral helium atoms enhance the damping in certain temperature interval, where the ratio of neutral helium and neutral hydrogen atoms is increased. Therefore, the height-dependence of ionization degrees of hydrogen and helium may influence the damping rate of Alfv\'en waves.}{We aim to study the effect of neutral helium in the damping of Alfv\'en waves in stratified partially ionized plasma of the solar chromosphere.}{We consider a magnetic flux tube, which is expanded up to 1000 km height and then becomes vertical due to merging with neighboring tubes, and study the dynamics of linear torsional Alfv\'en waves in the presence of neutral hydrogen and neutral helium atoms. We start with three-fluid description of plasma and consequently derive single-fluid magnetohydrodynamic (MHD) equations for torsional Alfv\'en waves. Thin flux tube approximation allows to obtain the dispersion relation of the waves in the lower part of tubes, while the spatial dependence of steady-state Alfv\'en waves is governed by Bessel type equation in the upper part of tubes.
}{Consecutive derivation of single-fluid MHD equations results in a new Cowling diffusion coefficient in the presence of neutral helium which is different from previously used one. We found that shorter-period ($<$ 5 s) torsional Alfv\'en waves damp quickly in the chromospheric network due to ion-neutral collision. On the other hand, longer-period ($>$ 5 s) waves do not reach the transition region as they become evanescent at lower heights in the network cores.}{Propagation of torsional Alfv\'en waves through the chromosphere into the solar corona should be considered with caution: low-frequency waves are evanescent due to the stratification, while high-frequency waves are damped due to ion neutral collisions.}

\keywords{Sun: atmosphere -- Sun: oscillations}

\titlerunning{Alfv\'en waves in partially ionized plasma}

\authorrunning{Zaqarashvili et al.}

\maketitle

\section{Introduction}

Alfv\'en waves play an important role in the dynamics of the solar atmosphere. Photospheric motions may excite the waves, which then propagate upwards along anchored magnetic field and transport the energy into upper layers, where they may deposit the energy back leading to the chromospheric/coronal heating and/or the acceleration of solar wind particles. Photospheric magnetic field is concentrated in flux tubes, which are very dynamic and may continuously change shape and/or merge due to granular motions. Nevertheless, in a crude representation useful for theoretical modeling, the magnetic field can be approximated as axis-symmetric tubes, therefore the excited pure Alfv\'en waves are axis-symmetric i.e. they are torsional Alfv\'en waves. Considering a cylindrically symmetric flux tube, torsional waves correspond to the azimuthal wavenumber set to $m=0$ in the standard notation. On the other hand, granular buffeting may excite transverse magnetohydrodynamic (MHD) kink waves in the tubes, which then can be transformed into linearly polarized Alfv\'en waves in upper layer due to the expansion of tubes (Cranmer and van Ballegooijen \cite{Cranmer2005}). The transverse oscillations have been observed with both, imaging and spectroscopic observations (Kukhianidze et al. \cite{Kukhianidze2006}, Zaqarashvili et al. \cite{Zaqarashvili20072}, De Pontieu et al. \cite{De Pontieu2007}, interested reader may find the detailed review of oscillations in Zaqarashvili \& Erd{\'e}lyi \cite{Zaqarashvili2009}).

Torsional Alfv\'en waves do not lead to the displacement of magnetic tube axis, therefore they can be observed only with spectroscopic observations as periodic variation of spectral line width (Zaqarashvili \cite{Zaqarashvili2003}). Observations of torsional Alfv\'en waves were reported recently in chromospheric spectral lines (Jess et al. \cite{Jess2009}, De Pontieu et al. \cite{De Pontieu2012}). Upward propagating undamped Alfv\'en waves can be also observed in the solar corona as an increase of non-thermal broadening of coronal spectral lines with height (Hassler et al. \cite{Hassler1990}).

The dynamics of Alfv\'en waves in the chromosphere/corona and their role in plasma heating are well studied (Hollweg \cite{Hollweg1981,Hollweg1984}, Copil et al. \cite{Copil2008}, Antolin and Shibata \cite{Antolin2010}, Vasheghani Farahani et al. \cite{Vasheghani2010,Vasheghani2011}, Morton et al. \cite{Morton2011}). They can be excited due to the vortex motion at the photospheric level (Fedun et al. \cite{Fedun2011a}). In the case of inhomogeneous plasma Alfv\'en waves may also be excited due to resonances with other wave modes (see, e.g., Soler et al. \cite{Soler2012}). Torsional Alfv\'en waves can be used as supplementary tool for solar magnetoseismology (Zaqarashvili and Murawski \cite{Zaqarashvili20071}, Fedun et al. \cite{Fedun2011b}, Verth et al. \cite{Verth2010}).

In the solar chromosphere and prominences plasma is only partially ionized which leads to the damping of Alfv\'en waves due to collision between ions and neutral hydrogen atoms (De Pontieu et al. \cite{De Pontieu2001}, Khodachenko et al. \cite{Khodachenko2004}, Leake et al. \cite{Leake2005}, Forteza et al. \cite{Forteza2007}, Soler et al. \cite{Soler2009}, Carbonell et al. \cite{Carbonell2010}, Singh and Krishan \cite{Singh}). Upward propagating Alfv\'en waves may also drive spicules through ion-neutral collisions (Haerendel \cite{Haerendel1992}, James \& Erd\'elyi \cite{James2002}, Erd\'elyi \& James \cite{Erdelyi2004}).

Single-fluid MHD description is a good approximation for low-frequency waves, but it fails when the wave frequency approaches to ion-neutral collision frequency, and consequently the multi-fluid MHD description should be used. It was shown by Zaqarashvili et al. (\cite{Zaqarashvili2011a}) that the damping rate is maximal for the waves those frequency is near ion-neutral collision frequency i.e. higher and lower frequency waves have less damping rates.

Beside the neutral hydrogen atoms, the solar plasma may contain significant amount of neutral helium atoms, which may enhance the damping of Alfv\'en waves. Soler et al. (\cite{Soler2010}) suggested that the neutral helium has no significant influence on the damping rate in the prominence cores with 8000 K temperature. On the other hand, Zaqarashvili et al. (\cite{Zaqarashvili2011b}) showed that the neutral helium may significantly enhance the damping of Alfv\'en waves in the temperature interval of 10000-40000 K, where the ratio of neutral helium and neutral hydrogen atoms is increased. This means that the helium atoms can be important in upper chromosphere, spicules and prominence-corona transition regions. The calculation of neutral helium effects in Zaqarashvili et al. (\cite{Zaqarashvili2011b}) was done in homogeneous medium with uniform magnetic field and ionization degree. On the other hand, the ion and neutral atom number densities and consequently ionization degree are significantly changed with height in the chromosphere, which may influence the damping of Alfv\'en waves due to ion-neutral collisions.

Stratification may significantly influence the dynamics of waves leading to their reflection at transition region, wave mutual transformation and/or evanescence. It introduces a cut-off frequency for Alfv\'en waves in fully ionized plasma, i.e. waves with lower frequency than the cut-off value are evanescent in the solar atmosphere (Musielak et al. \cite{Musielak1995}, Murawski \& Musielak \cite{Murawski2010}).

In this paper, we study the effects of stratification and neutral helium atoms on the propagation/damping of Alfv\'en waves in the solar chromosphere. We consider a magnetic flux tube, which is expanded up to 1000 km height and then becomes vertical due to merging with neighboring tubes. Consequently, we consider torsional Alfv\'en waves, which are only pure incompressible waves in tubes.

\section{Linear torsional Alfv\'en waves}

We consider a vertical magnetic flux tube embedded in the stratified solar atmosphere. We use cylindrical coordinate system ($r, {\theta}, z$) and suppose that the unperturbed magnetic field, $\vec B_0$, is untwisted i.e. $B_{0\theta}=0$. Plasma consists in electrons, protons ($\mathrm {H}^+$), singly ionized helium ($\mathrm {He^+}$), neutral hydrogen ($\mathrm {H}$) and neutral helium ($\mathrm {He}$) atoms. We consider the Alfv\'en waves polarized in $\theta$ direction i.e. the only non-zero components of the perturbations are $v_{\theta}$ and $b_{\theta}$. Therefore, the waves are torsional Alfv\'en waves. We start with the three-fluid approach for partially ionized hydrogen-helium plasma, when one component is ion+electron gas and other two components are neutral hydrogen and neutral helium gasses (Zaqarashvili et al. \cite{Zaqarashvili2011b}). Then, the linear Alfv\'en waves are governed by the equations (magnetic diffusion due to ion-electron collision is neglected):
\begin{equation}\label{Alfven-vl}
{{\partial }\over {\partial t}}(rv_{\theta})=
{\vec B_0{\cdot}\nabla\over {4 \pi
\rho_{i}}}(rb_{\theta}) -{{\alpha_{{\mathrm {H}}}+\alpha_{{\mathrm {He}}}}\over \rho_{i}}(rv_{\theta})+{{\alpha_{{\mathrm {H}}}}\over \rho_{i}}(rv_{\mathrm {H}\theta})+{{\alpha_{{\mathrm {He}}}}\over \rho_{i}}(rv_{\mathrm {He}\theta}),
\end{equation}
\begin{equation}\label{Alfven-uHl}
{{\partial }\over {\partial t}}(rv_{\mathrm {H}\theta})={{\alpha_{{\mathrm {H}}}}\over \rho_{H0}}(rv_{\theta})-{{\alpha_{{\mathrm {H}}}+\alpha_{{\mathrm {HeH}}}}\over \rho_{{\mathrm {H}}0}}(rv_{\mathrm {H}\theta})+{{\alpha_{{\mathrm {HeH}}}}\over \rho_{H0}}(rv_{\mathrm {He}\theta}),
\end{equation}
\begin{equation}\label{Alfven-uHel}
{{\partial }\over {\partial t}}(rv_{\mathrm {He}\theta})={{\alpha_{{\mathrm {He}}}}\over \rho_{{\mathrm {He}}0}}(rv_{\theta})-{{\alpha_{{\mathrm {He}}}+\alpha_{{\mathrm {HeH}}}}\over \rho_{{\mathrm {He}}0}}(rv_{\mathrm {He}\theta})+{{\alpha_{{\mathrm {HeH}}}}\over \rho_{He0}}(rv_{\mathrm {H}\theta}),
\end{equation}
\begin{equation}\label{Alfven-by}
{{\partial b_{\theta}}\over {\partial t}}=r(\vec B_0{\cdot}\nabla)\left ({{v_{\theta}}\over {r}}\right ),
\end{equation}
where  $v_{\theta}$ ($v_{\mathrm {H}\theta}$, $v_{\mathrm {He}\theta}$) are the perturbations of ion (neutral hydrogen, neutral helium) velocity, $b_{\theta}$ is the perturbation of the magnetic field, and $\rho_{i}$ ($\rho_{{\mathrm {H}}0}$, $\rho_{{\mathrm {He}}0}$) is the unperturbed ion (neutral hydrogen, neutral helium) density. Here we use the definitions
\begin{equation}\label{a}
\alpha_{{\mathrm {H}}}=\alpha_{{\mathrm {H^+H}}}+\alpha_{{\mathrm {He^+H}}},\,\, \alpha_{{\mathrm {He}}}=\alpha_{{\mathrm {H^+He}}}+\alpha_{{\mathrm {He^+He}}},
\end{equation}
where $\alpha$ denotes the coefficient of friction between different sort of species. The collision between electrons and neutral atoms is also neglected as it has much less effect in the damping of MHD waves than the collision between ions and neutrals. In the present model the motions of torsional Alfv\'en waves are normal to the direction of gravity, so that gravity does not explicitly appear in the equations. The effect of gravity is indirectly present in the equations through the stratification of plasma density.

\begin{figure}[t]
\vspace*{1mm}
\begin{center}
\includegraphics[width=9.0cm]{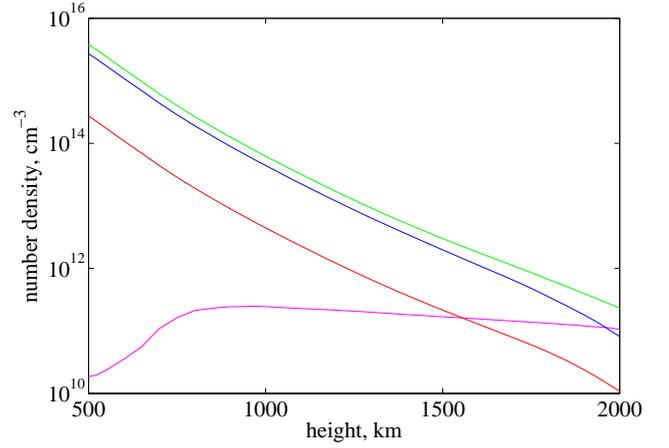}
\end{center}
\caption{Height dependence of atmospheric parameters according to FAL93-F model (Fontenla et al. \cite{Fontenla1993}): magenta, red and blue lines correspond to proton, neutral helium and neutral hydrogen number densities, respectively. Green line is the total (proton+neutral hydrogen+neutral helium) number density.}
\end{figure}

The coefficient of friction between ions and neutrals (in the case of same temperature) is calculated as (Braginskii \cite{Braginskii1965})
\begin{equation}\label{in}
{\alpha_{in}}= n_i n_n m_{in}\sigma_{in}{4\over 3} \sqrt{{{8kT}\over {\pi m_{in}}}},
\end{equation}
where $T$ is the plasma temperature, $m_i$ ($m_n$) the ion (neutral atom) mass, $m_{in}=m_i m_n /(m_i+m_n)$ is reduced mass, $n_i$ ($n_n$) \textbf{is} the ion (neutral atom) number density, $\sigma_{in}=\pi (r_i+r_n)^2 \approx \pi
r^2_n$ is ion-neutral collision cross section for elastic collisions from Braginskii (\cite{Braginskii1965}) (mean atomic cross section is $\pi r^2_n=8.7974\times 10^{-17}$ cm$^{2}$), and $k=1.38\times 10^{-16}$ erg K$^{-1}$ is the Boltzmann constant.

From here we adopt the single-fluid approximation. To justify whether or not the single-fluid approximation is valid in the solar atmosphere let us estimate the ion-neutral collision frequency.
Mean collision frequency between ions and neutrals can be calculated as (Zaqarashvili et al. \cite{Zaqarashvili2011b})
\begin{equation}\label{in_t}
\nu_{in}={\alpha_{in}}\left ({{1}\over {m_in_i}}+{{1}\over {m_nn_n}}\right ).
\end{equation}
The collision frequency is very high in the photosphere and decreases upwards. For example, The collision frequency between protons and neutral hydrogen can be estimated as $8.6 \cdot 10^{6}$ Hz, $6.2 \cdot 10^{3}$ Hz and 24 Hz, at $z=0$, $z=900$ km and $z=1900$ km correspondingly (ion and neutral atom number densities were taken from Fontenla et al. \cite{Fontenla1993}, model-F). Therefore, the Alfv\'en waves with periods $>$ 1 s can be studied in the single-fluid approach.

To obtain the governing equations in the single fluid approximation, we consider the total density
\begin{equation}\label{rho}
\rho=\rho_i+\rho_H+\rho_{He}
\end{equation}
and velocity of center of mass
\begin{equation}\label{V}
V_{\theta}={{\rho_i v_{\theta}+\rho_H v_{\mathrm {H}\theta} +\rho_{He}v_{\mathrm {He}\theta}}\over {\rho_i+\rho_H+\rho_{He}}}.
\end{equation}
We also consider the relative velocity of ion and neutral hydrogen as $w_{\mathrm {H}\theta}=v_{\theta}-v_{\mathrm {H}\theta}$ and the relative velocity of ion and neutral helium as $w_{\mathrm {He}\theta}=v_{\theta}-v_{\mathrm {He}\theta}$. Then one may find that
\begin{equation}\label{v_y}
v_{\theta}=V_{\theta}+\xi_H w_{\mathrm {H}\theta} +\xi_{He} w_{\mathrm {He}\theta},
\end{equation}
where $\xi_H=\rho_H/\rho$ and $\xi_{He}=\rho_{He}/\rho$.

Consecutive substraction of Eq. (\ref{Alfven-vl}) and Eq. (\ref{Alfven-uHl}), Eq. (\ref{Alfven-vl}) and Eq. (\ref{Alfven-uHel}), Eq. (\ref{Alfven-uHl}) and Eq. (\ref{Alfven-uHel}) and neglecting the inertial terms for relative velocities lead to the equations
\begin{equation}\label{w_H1}
rw_{\mathrm {H}\theta}=\left [{{\alpha_{He}}\over {\alpha}}\xi_H+{{\alpha_{HeH}}\over {\alpha}}(\xi_H+\xi_{He})\right ]{\vec B_0{\cdot}\nabla\over {4 \pi}}(rb_{\theta}),
\end{equation}
\begin{equation}\label{w_He1}
rw_{He}=\left [{{\alpha_{H}}\over {\alpha}}\xi_{He}+{{\alpha_{HeH}}\over {\alpha}}(\xi_H+\xi_{He})\right ]{\vec B_0{\cdot}\nabla\over {4 \pi}}(rb_{\theta}),
\end{equation}
where $\alpha=\alpha_{H}\alpha_{He}+\alpha_{H}\alpha_{HeH}+\alpha_{He}\alpha_{HeH}$. The neglecting of the inertial terms is justified for waves those frequency is less than ion-neutral collision frequency. This is the key step to transform the multi-fluid equations into the single fluid description.

Then the sum of Eqs. (\ref{Alfven-vl})-(\ref{Alfven-uHel}) and using Eqs. (\ref{Alfven-by}), (\ref{v_y}), (\ref{w_H1}) and (\ref{w_He1}) lead to the single-fluid equations
\begin{equation}\label{eqs1}
{{\partial }\over {\partial t}}(rV_{\theta})=
{\vec B_0{\cdot}\nabla\over {4 \pi
\rho(z)}}(rb_{\theta}),
\end{equation}
\begin{equation}\label{eqs2}
{{\partial b_{\theta}}\over {\partial t}}=r(\vec B_0{\cdot}\nabla)\left ({{V_{\theta}}\over {r}}\right )
+r(\vec B_0{\cdot}\nabla)\left [ {{\eta_c(z)}\over {B^2_0}} {\vec B_0{\cdot}\nabla\over {r^2}}(rb_{\theta})\right ],
\end{equation}
where
\begin{equation}\label{eqs3}
\eta_c={B^2_0\over {4 \pi}}{{\alpha_{He}\xi^2_{H} +\alpha_{H}\xi^2_{He}+\alpha_{HeH}(\xi_{H}+\xi_{He})^2}\over {\alpha_{H}\alpha_{He}+\alpha_{H}\alpha_{HeH}+\alpha_{He}\alpha_{HeH}}}
\end{equation}
is the coefficient of Cowling diffusion. The first two terms in the nominator of Eq. (\ref{eqs3}) are due the frictions of neutral hydrogen and neutral helium separately with regards to ions, while the last term is the friction of neutral hydrogen+helium fluid with regards to ions. Therefore, the last term is important in weakly ionized plasmas, while the first two terms are important for relatively high value of ionization degree. Soler et al. (\cite{Soler2010}) used the following expression for Cowling diffusion in the presence of neutral helium
\begin{equation}\label{eqss}
\eta_c={B^2_0\over {4 \pi}}{{(\xi_{H}+\xi_{He})^2}\over {\alpha_H+\alpha_{He}}}
\end{equation}
which can be obtained from our expression (Eq. \ref{eqs3}) when $\alpha_H,\alpha_{He}\ll\alpha_{HeH}$ i.e. when plasma is only weakly ionized. This is so because Soler et al. (\cite{Soler2010}) considered in their derivation that both neutral hydrogen and neutral helium have the same velocity. That is equivalent to take $\alpha_{HeH} \to \infty$. Therefore the present description is more general than that of Soler et al. (\cite{Soler2010}). The expression used by Soler et al. (\cite{Soler2010}) is only approximately valid for the photosphere and lower chromosphere, while in the upper chromosphere, spicules and prominences the general expression Eq. (\ref{eqs3}) should be used.

\begin{figure}[t]
\vspace*{1mm}
\begin{center}
\includegraphics[width=9.0cm]{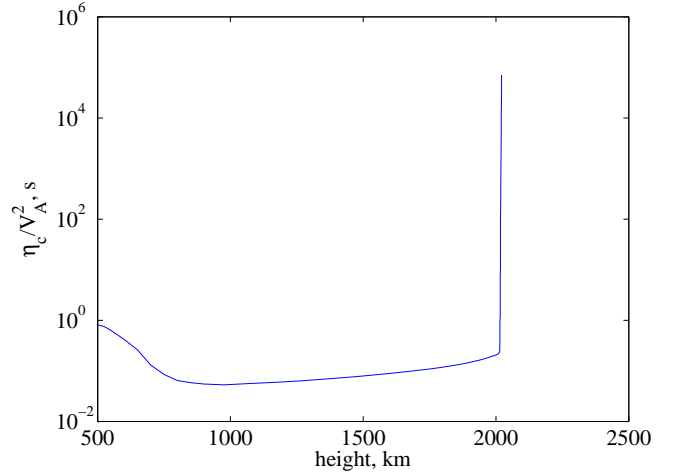}
\end{center}
\caption{The ratio of Cowling diffusion coefficient and Alfv\'en speed square, $\eta_c/V^2_A$, vs height according to FAL93-F model (Fontenla et al. \cite{Fontenla1993}).
}
\end{figure}

Let us now consider a new coordinate $s$, which represents a longitudinal coordinate along the unperturbed magnetic field, $\vec B_0$, then Eqs. (\ref{eqs1})-(\ref{eqs2}) can be combined into the single equation in terms of this new coordinate, namely
$$
{{\partial^2 {U_{\theta}}}\over {\partial t^2}}={{B_{0s}}\over {4\pi \rho r^2}} {{\partial }\over {\partial s}}\left [r^2 B_{0s} {{\partial {U_{\theta}}}\over {\partial s}}\right ]+
$$
\begin{equation}\label{eqs4}
+{{B_{0s}}\over {4\pi \rho r^2}} {{\partial }\over {\partial s}}\left [r^2 B_{0s} {{\partial }\over {\partial s}} \left ({{4\pi \rho \eta_c}\over {B^2_{0s}}}{{\partial {U_{\theta}}}\over {\partial t}} \right )\right ],
\end{equation}
where $U_{\theta}=V_{\theta}/r$. Note that $r$ is a function of $s$. This equation is simplified near the axis of symmetry, where $B_{0s}(s)r^2(s) \approx constant$ (Hollweg \cite{Hollweg1981}), which is the conservation of magnetic flux near the tube axis, where field lines are almost perpendicular to the tube cross-section. Then this equation may be written as
\begin{equation}\label{eqs4}
{{\partial^2 {U_{\theta}}}\over {\partial t^2}}=V^2_A(s) {{\partial^2 }\over {\partial s^2}}\left [\left (1+ {{\eta_c(s)}\over {V^2_A(s)}} {{\partial }\over {\partial t}}\right ){U_{\theta}}\right ],
\end{equation}
where \begin{equation}\label{alfven-speed}
V_A= {B_{0s}(s)\over {\sqrt{4 \pi \rho(s)}}}
\end{equation}
is the Alfv\'en speed.

Solar atmospheric parameters depend on the height due to gravity. Plasma ionization degree is also a function of altitude due the increase of temperature with height. The plasma is only weakly ionized in the photosphere/lower chromosphere. The increase in temperature with height leads to the ionization of hydrogen and helium atoms, which become almost fully ionized in the solar corona, so the transition occurs near the region of sharp temperature rise.

\begin{figure}[t]
\vspace*{1mm}
\begin{center}
\includegraphics[width=8.5cm]{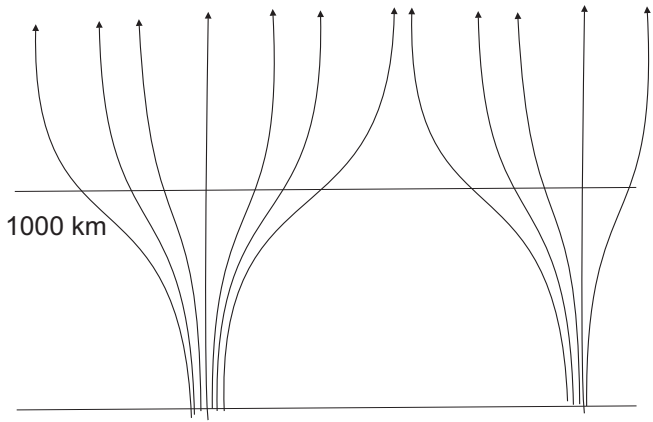}
\end{center}
\caption{Vertical magnetic flux tube model: below 1000 km the Alfv\'en speed is constant due to the thin flux tube approximation and above 1000 km
height the Alfv\'en speed increases exponentially.
}
\end{figure}

Figure 1 shows the density of different species vs height according to the FAL93-F model (Fontenla et al. \cite{Fontenla1993}). This model includes the dependence of ionization degree on heights for both hydrogen and helium. The neutral atom number densities are much higher than the ion number densities at the lower heights, but become comparable near $\sim$ 1900 km, which corresponds to the temperature of about 10000 K. It is seen that the neutral atoms are more stratified than protons between 500 and 2000 km heights: the neutral hydrogen, neutral helium and total number densities have scale height about 180 km, while the scale height of protons is much larger. The total number density is similar to the neutral hydrogen number density until $\sim$1500 km. Above this height they start to diverge due to the increase of ionization degree.

Fourier analysis of Eq. (\ref{eqs4}) with $\exp[-i \omega t]$, where $\omega$ is the wave frequency, gives the equation
\begin{equation}\label{strat0}
V^2_A(s) {{\partial^2 }\over {\partial s^2}}\left [\left (1- i{{\omega \eta_c(s)}\over {V^2_A(s)}}\right ){U_{\theta}}\right ]+\omega^2 {U_{\theta}}=0,
\end{equation}
which governs the spatial dependence of steady-state torsional Alfv\'en waves. The type of the equation depends on the dependence of $V_A$ and $\eta_c$ on $s$.

Very important parameter in Eq. (\ref{strat0}) is the ratio $\eta_c/V^2_A$, which does not depend on the magnetic field structure with height as $B^2_{0s}$ is canceled. We plot the ratio vs height according to VAL93-F model on Fig. 2. It is clearly seen that the ratio is almost constant along whole chromosphere and abruptly increases near the transition region.  Therefore, in order to find analytical solutions to Eq. (\ref{strat0}) we consider the ratio as a constant along $s$ in the chromosphere
\begin{equation}\label{ratio}
{\eta_c(s)\over V^2_A(s)}\approx const.
\end{equation}
From Figure~2 we see that this approximation seems to be good in the upper part of the chromosphere, i.e.,  between 800-2000 km, while it may be not so accurate in the lower part of the chromosphere, i.e., in the interval between 500-800 km. Nevertheless, we use this approximation along whole chromosphere for convenience.

Plasma $\beta$ (=$8 \pi p/B^2_{0s}$) is constant with height in the isothermal atmosphere in thin vertical magnetic flux tubes, when temperatures inside and outside the tubes are same (Roberts \cite{Roberts2004}). In this case, the Alfv\'en speed is also constant with height yielding $V_A(s)=V_{A0}=const$. The constancy of the Alfv\'en speed in thin tubes is caused due to the compensation of density variation by the magnetic field strength variation with height. The thin flux tube approximation is valid up to 1000 km above the photosphere (Hasan et al. \cite{Hasan2003}, Cranmer and van Ballegooijen \cite{Cranmer2005}). Above this height, the magnetic tubes are thick and they probably merge with neighboring tubes producing almost vertical magnetic field (Fig. 3). Then, the Alfv\'en speed may vary with height due to the decrease of density, while the magnetic field is constant. According to this model, we consider the vertical magnetic tube which consists in two different parts: the lower part with constant Alfv\'en speed and the upper part with exponentially increasing Alfv\'en speed. Then we study the torsional Alfv\'en waves in the two parts of magnetic flux tube separately.

\begin{figure}
\begin{center}
\includegraphics[width=9cm]{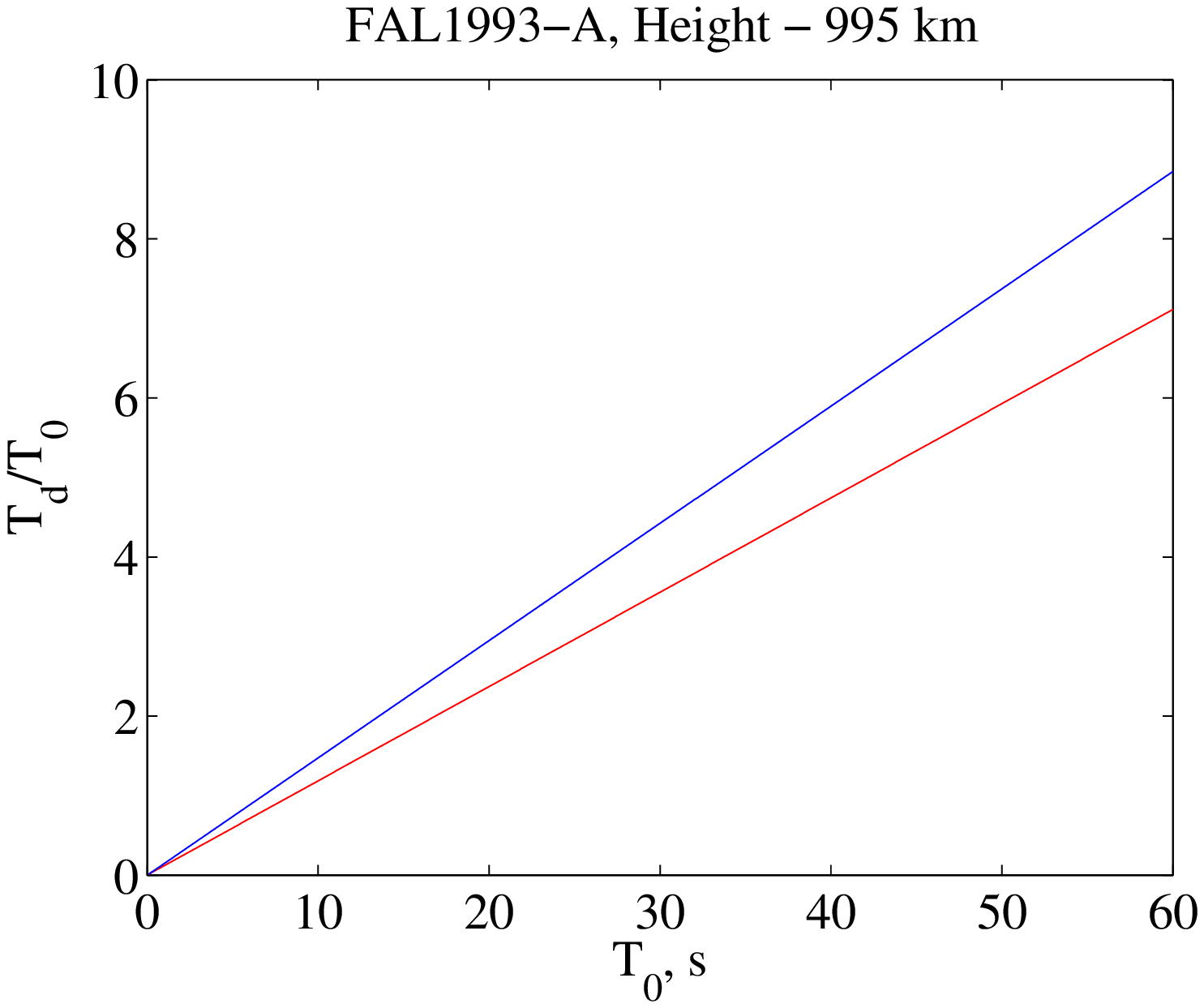}
\includegraphics[width=9cm]{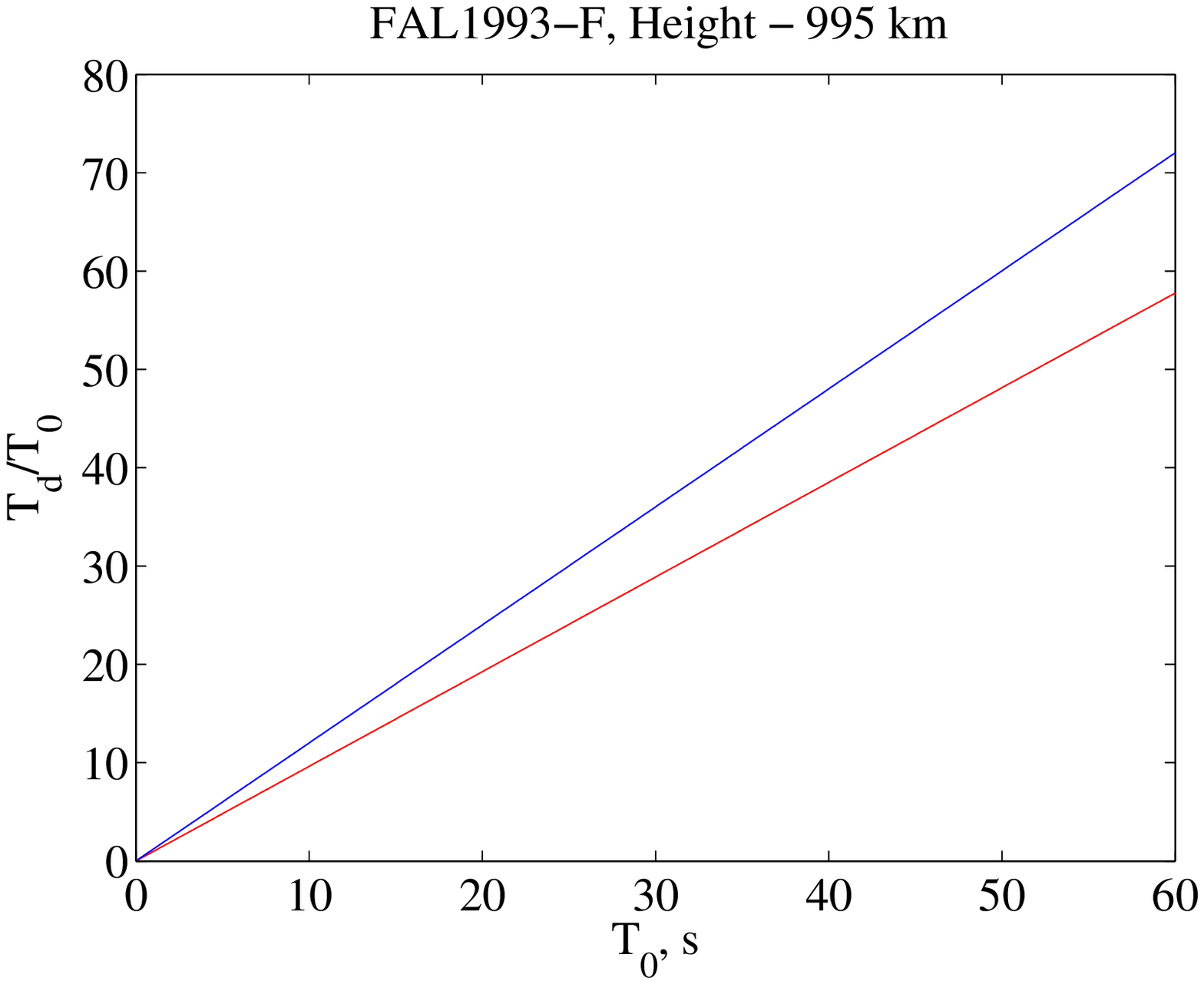}
\includegraphics[width=9cm]{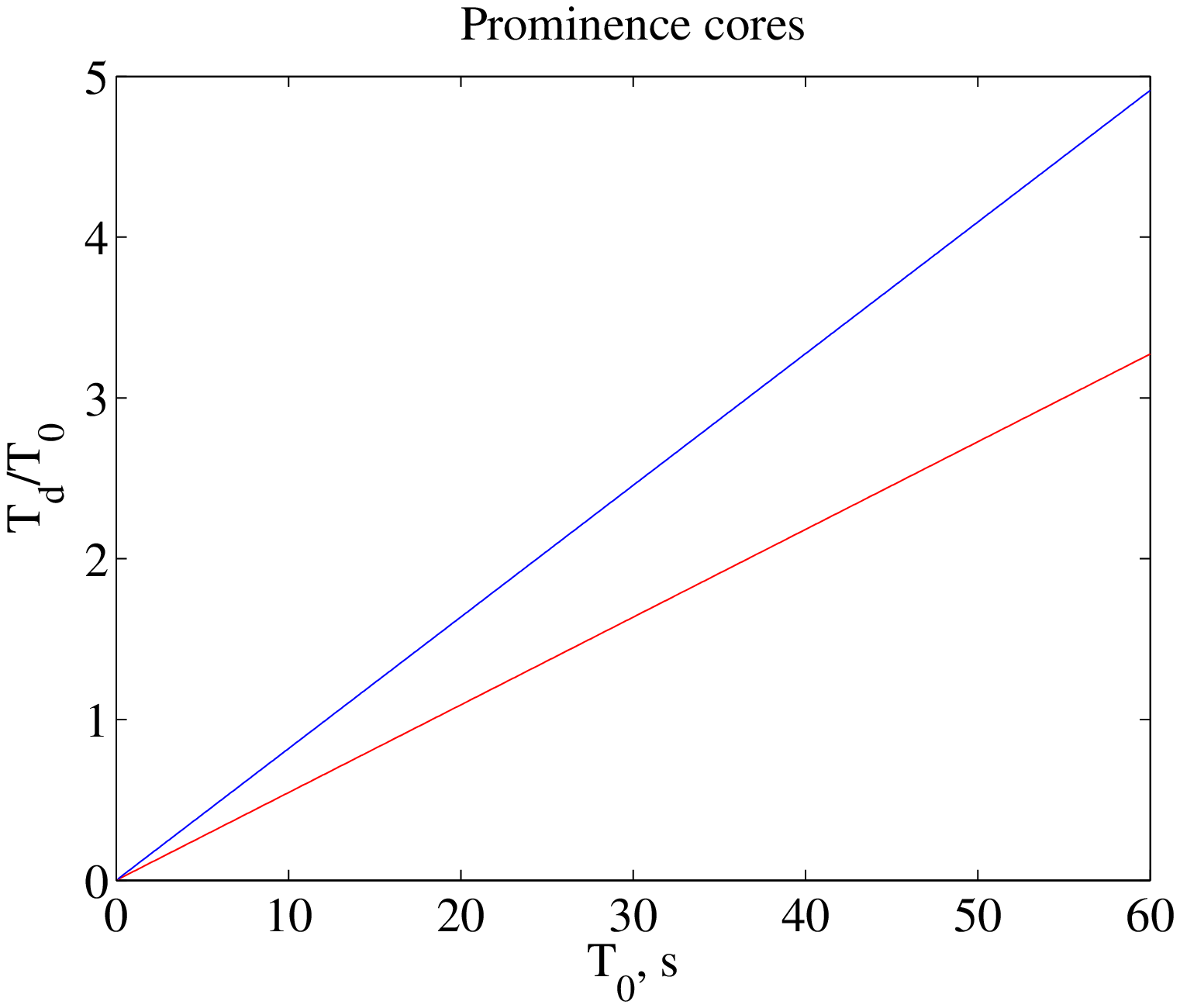}
\end{center}
\caption{Normalized damping time of torsional Alfv\'en waves ($T_d/T_0$) vs wave period ($T_0$) for three different situations from top to bottom: faint cell center area (FAL93-A), bright network (FAL93-F) and prominence core. The blue line indicates the wave damping due to collision of ions with neutral hydrogen atoms only. While the red line indicates the wave damping due to collision of ions with neutral hydrogen and neutral helium atoms. }
\end{figure}

\section{Lower chromosphere: thin flux tube approximation}

In the lower part of magnetic flux tube, the thin tube approximation is used which yields $V_A(s)=const$ (see previous section). Note that the thin tube approximation is still compatible with a magnetic flux tube expanding with height. The only restriction is that the wavelength remains much longer than the tube radius. This case was also studied by Soler et al. (\cite{Soler2009})  for the case of torsional Alfv\'en waves in prominence threads. The constancy of Alfv\'en speed means that the coefficient of Cowling diffusion is also constant. Then the coefficients of Eq. (\ref{strat0}) do not depend on $s$, therefore Fourier analysis with $exp[i(k_s s -\omega t)]$ gives the dispersion relation
\begin{equation}\label{disp-he}
\omega^2+i\eta_c k^2_s \omega -k^2_s V^2_A=0,
\end{equation}
which has two complex solutions
\begin{equation}\label{solution}
\omega= \pm k_s V_A \sqrt{1-{{\eta^2_c k^2_s}\over {4 V^2_A}}} -i {{\eta_c k^2_s}\over {2}}.
\end{equation}
The dispersion relation and its solutions agree with Equations (33) and (34) of Soler et al. (\cite{Soler2009}).
Real part of this expression gives the wave frequency, which shows that there is the cut-off wavenumber due to Cowling's diffusion, $k_s=2V_A/\eta_c$. The cut-off wave number due to ion-neutral collision was found in different situations of the solar prominences (Forteza et al. \cite{Forteza2007}, Soler et al. \cite{Soler2009}, Barc\'elo et al. \cite{Barcelo2011}). It is the result of approximation when one transforms multi-fluid equations into the single-fluid MHD and therefore it has no physical ground (Zaqarashvili et al. \cite{Zaqarashvili2012}).

On the other hand, the imaginary part of this expression gives the normalized damping rate as
\begin{equation}\label{damping}
|{\tilde \omega_i}|= \left |{{\omega_i}\over {k_s V_A}}\right |= {1\over 2} {{k_s V_A}\over {\rho}}{{\alpha_{He}\rho^2_{H} +\alpha_{H}\rho^2_{He}+\alpha_{HeH}(\rho_{H}+\rho_{He})^2}\over {\alpha_{H}\alpha_{He}+\alpha_{H}\alpha_{HeH}+\alpha_{He}\alpha_{HeH}}},
\end{equation}
where we have used the full expression of Cowling's coefficient, $\eta_c$. The plasma is almost neutral in the low chromosphere, therefore $\alpha_{HeH}\gg \alpha_{H},\alpha_{He}$. This leads to the normalized damping rate
\begin{equation}\label{damping1}
|{\tilde \omega_i}|\approx{1\over 2} {{k_s V_A}\over {\rho}} \left [{{(\rho_{H}+\rho_{He})^2}\over {\alpha_{H}+\alpha_{He}}}\right ].
\end{equation}
This expression was used by Soler et al. (2010) to calculate the effect of neutral helium in prominence cores.

On the other hand, when the number density of neutral atoms is much less than ion number density, then $\alpha_{HeH}\ll \alpha_{H},\alpha_{He}$ and the normalized damping rate is
\begin{equation}\label{damping2}
|{\tilde \omega_i}|\approx{1\over 2} {{k_s V_A}\over {\rho}} \left [{{\rho^2_{H}}\over {\alpha_{H}}}+{{\rho^2_{He}}\over {\alpha_{He}}}\right ].
\end{equation}

When ion number density is of the same order as the number density of neutral atoms (like in spicules or in prominence cores), then the general expression of damping rate (Eq. \ref{damping}) should be used. The expression Eq. \ref{damping} may significantly change the contribution of neutral helium atoms.

%
%\begin{rob2}
%Please define the period, $T_0$, and the damping time, $T_{\rm d}$, used in Fig. 4.
%\end{rob2}

Fig. 4 shows the normalized damping time ($T_{\rm d}/T_0=1/\tilde \omega_i$) vs the period of Alfv\'en waves ($T_0=2\pi/k_s V_A$) for three different areas: a) FAL93-A corresponding to faint cell center area, b) FAL93-F corresponding to bright area of chromospheric network, and c) prominence cores. We use the following parameters: a) $n_i$=3.26${\times}$10$^{10}$ cm$^{-3}$, $n_H$=3.01${\times}$10$^{13}$ cm$^{-3}$, $n_{He}$=3.01${\times}$10$^{12}$ cm$^{-3}$ taken from FAL93-A at {\bf 975} km height corresponding to $\approx$ 5480 K temperature; b) $n_i$=2.49${\times}$10$^{11}$ cm$^{-3}$, $n_H$=5.24${\times}$10$^{13}$ cm$^{-3}$, $n_{He}$=5.26${\times}$10$^{12}$ cm$^{-3}$ was taken from FAL93-F at 975 km height corresponding to 6180 K temperature and
c) $n_i$=10$^{10}$ cm$^{-3}$, $n_H$=2${\times}$10$^{10}$ cm$^{-3}$, $n_{He}$=2${\times}$10$^{9}$ cm$^{-3}$ was taken for prominence cores corresponding to 8000 K temperature. It is clearly seen that the presence of neutral helium atoms significantly enhances the damping of Alfv\'en waves. It is also seen that the damping of torsional Alfv\'en waves is more efficient in bright cell center area in the chromosphere and in prominence cores, where the waves are damped over few wave periods. On the other hand, the damping of torsional Alfv\'en waves due to ion-neutral collision is inefficient in bright chromospheric network center. The clear difference of damping in these two areas can be understood in terms of ion-neutral collision frequency. In the absence of neutral helium atoms, Eq. \ref{damping} can be rewritten as (Zaqarashvili et al. \cite{Zaqarashvili2011b})
\begin{equation}\label{damping5}
|{\tilde \omega_i}|\approx{1\over 2} {{k_s V_A}\over {\nu_{iH}}}{{\rho_{H}}\over {\rho_{i}}},
\end{equation}
which clearly indicates that the normalized damping rate depends on the ratio of Alfv\'en frequency over ion-neutral hydrogen collision
frequency ($\nu_{iH}$) and the ratio of neutral and ion fluid densities. For $\rho_i \ll \rho_H$, which is the case at 975 km height, ion-neutral hydrogen collision frequency is proportional to neutral hydrogen density, $\nu_{iH}\sim \rho_H$. This means that the normalized damping rate is inversely proportional to ion number density, $\rho_{i}$. Proton number density is almost 10 times smaller in the cell center than in the network center, which leads to almost 10 times difference in damping (Fig. 4).

%In section 2.1 (constant Alfv\'en speed) you fix a real $k_s$ and solve the dispersion relation to obtain a complex frequency. You compute the damping time from the imaginary part of $\omega$. This means that you are actually studying the temporal damping of standing waves. This is OK to compare your expressions with previous papers.

%However, in the section 2.2 (increasing Alfv\'en speed) you study the propagation of the waves for a fixed and real frequency, so that you study the spatial behaviour.  In my opinion, for consistency we should study the same situation in both sections. Maybe, it would be convenient to study the spatial damping of propagating waves in section 2.1 also, meaning that the dispersion relation should be solved for the complex wavenumber, $k_s$, assuming a fixed and real $\omega$. Later the damping length can be computed from the imaginary part of $k_s$.

In the first part of this section we have studied the temporal damping of Alfv\'en waves, i.e., we have set a real $k_s$ and solved the dispersion relation to obtain the complex $\omega$. In the following paragraphs we study the spatial damping. This means that the dispersion relation (Eq. \ref{disp-he}) is now solved for the complex wavenumber, $k_s$, assuming a fixed and real $\omega$. The solution of the dispersion relation for the square of $k_s$ is
\begin{equation}\label{ks}
k_s^2 = \frac{\omega^2}{V_A^2 - i \omega \eta_c}.
\end{equation}
We use this expression to find the real and imaginary parts of $k_s$, assuming $k_s = k_r + i k_i$. After some algebraic manipulations we can obtain the expressions for $k_r^2$ and $k_i$, namely
\begin{eqnarray}
k_r^2 &=&\frac{\omega^2 V_A^2}{2(V_A^4 + \omega^2 \eta_c^2)} \pm \frac{\omega^2 V_A^2}{2(V_A^4 + \omega^2 \eta_c^2)}  \left( 1 + \frac{\eta_c^2 \omega^2}{V_A^4} \right)^{1/2}, \label{ksr2} \\
k_i &=& \frac{1}{k_r} \frac{\eta_c \omega^3}{2(V_A^4+\omega^2 \eta_c^2)}.
\end{eqnarray}
The expression of $k_r$ is computed from Eq.~(\ref{ksr2}) as $k_r = \pm \sqrt{k_r^2}$. Note that the $\pm$ sign in front of the second term in the expression of $k_r^2$ leads to 4 possible values of $k_r$. The two values of $k_r$ with $+$ sign in the expression of $k_r^2$ correspond to propagating waves, namely
\begin{equation}
k_r = \pm \sqrt{\frac{\omega^2 V_A^2}{2(V_A^4 + \omega^2 \eta_c^2)} + \frac{\omega^2 V_A^2}{2(V_A^4 + \omega^2 \eta_c^2)}  \left( 1 + \frac{\eta_c^2 \omega^2}{V_A^4} \right)^{1/2}}.
\end{equation}
where the $\pm$ sign in front of the square root refers to upward and downward waves, respectively.

\begin{figure}
\begin{center}
\includegraphics[width=9.5cm]{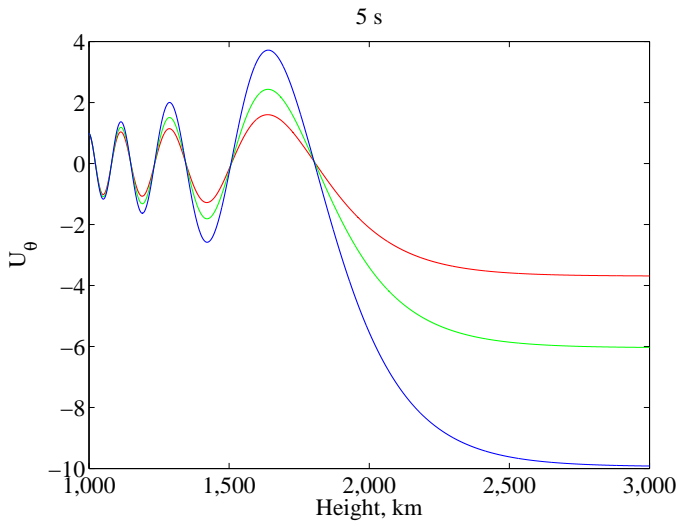}
\end{center}
\caption{Height dependence of steady state torsional Alfv\'en waves with period of 5 s. Green line corresponds to $H^{1}_0(\zeta)$ and $H^{2}_0(\zeta)$ from Eq. (\ref{strat}) in fully ionized plasma. Blue and Red lines correspond to $H^{1}_0(\zeta)$ and $H^{2}_0(\zeta)$, respectively, in partially ionized plasma.}
\end{figure}

To obtain more simplified analytical formulas let us consider the case of weak damping, i.e., $\frac{\eta_c^2 \omega^2}{V_A^4} \ll 1$. The expression of $k_r$ for propagating waves becomes
\begin{equation}
k_r \approx \pm \sqrt{\frac{\omega^2 V_A^2}{V_A^4 + \omega^2 \eta_c^2} \left( 1 + \frac{\omega^2 \eta_c^2}{4V_A^4}\right)}\approx \pm \frac{\omega}{V_A} \left( 1 - \frac{3}{8}\frac{\omega^2 \eta_c^2}{V_A^4} \right).
\end{equation}
We now take the $+$ sign corresponding to upward propagating waves and use this expression to compute the value of $k_i$ in the limit $\frac{\eta_c^2 \omega^2}{V_A^4} \ll 1$, namely
\begin{equation}
k_i \approx \frac{1}{2k_r} \frac{\eta_c \omega^3}{V_A^4} \approx \frac{\eta_c \omega^2}{2 V_A^3}.
\end{equation}
The corresponding normalized damping rate using the full expression of $\eta_c$ is
\begin{equation}
\tilde{k}_i = \frac{k_i}{\omega/V_A} = \frac{1}{2}\frac{\omega}{\rho}{{\alpha_{He}\rho^2_{H} +\alpha_{H}\rho^2_{He}+\alpha_{HeH}(\rho_{H}+\rho_{He})^2}\over {\alpha_{H}\alpha_{He}+\alpha_{H}\alpha_{HeH}+\alpha_{He}\alpha_{HeH}}},
\end{equation}
which is equivalent to Equation~(\ref{damping}) obtained for the temporal damping if we replace $\omega$ by $k_s V_A$. Therefore, both spatial and temporal damping rates of the Alfv\'en waves are equivalent.

On the other hand, the two values of $k_r$ corresponding to the $-$ sign in the expression of $k_r^2$ (Equation~(\ref{ksr2})) correspond to evanescent waves, namely
\begin{equation}
k_r = \pm \sqrt{\frac{\omega^2 V_A^2}{2(V_A^4 + \omega^2 \eta_c^2)} - \frac{\omega^2 V_A^2}{2(V_A^4 + \omega^2 \eta_c^2)}  \left( 1 + \frac{\eta_c^2 \omega^2}{V_A^4} \right)^{1/2}},
\end{equation}
which for weak damping, i.e., $\frac{\eta_c^2 \omega^2}{V_A^4} \ll 1$, becomes
\begin{equation}
k_r = \pm i \sqrt{\frac{\omega^4 \eta_c^2}{4V_A^2(V_A^4 + \omega^2 \eta_c^2)}}.
\end{equation}
This corresponds to evanescent perturbations that do not propagate from the location of the excitation, and so these solutions are not relevant for the present study.

%Some plots using these expressions could be added.

\section{Upper chromosphere: exponentially increasing Alfv\'en speed}

In the upper part of magnetic flux tube, where the neighboring tubes are merged, we assume that the Alfv\'en speed is exponentially increasing (note that the ratio of Cowling diffusion coefficient and Alfv\'en speed square is assumed to be constant again)
\begin{equation}\label{Al}
V_A(s)=V_{A0}\exp\left ({{s\over {2h}}}\right ),
\end{equation}
where $h$ is the scale height and $V_{A0}$ corresponds to the value of Alfv\'en speed at the height of 1000 km (i.e. $s=0$ also corresponds to the height of 1000 km). In this case, Eq. (\ref{strat0}) can be rewritten as
\begin{equation}\label{strat1}
{{\partial^2 {U_{\theta}}}\over {\partial s^2}}+\exp\left ({-{s\over h}}\right )k^2_{s0}{U_{\theta}}=0,
\end{equation}
where
\begin{equation}\label{ks0}
k_{s0}^2 = \frac{\omega^2}{V_{A0}^2 - i \omega \eta_{c0}}
\end{equation}
and $\eta_{c0}$ is the value of Cowling diffusion at the height of 1000 km. The solution of this equation is
\begin{equation}\label{strat}
{U_{\theta}}={U_{\theta}}(0)F_0(\zeta),
\end{equation}
where $F_0$ is Bessel, modified Bessel or Hankel functions of zero order and
$$
\zeta=2hk_{s0}\exp\left [{-{s\over {2h}}}\right ].
$$

\begin{figure}
\begin{center}
\includegraphics[width=8.3cm]{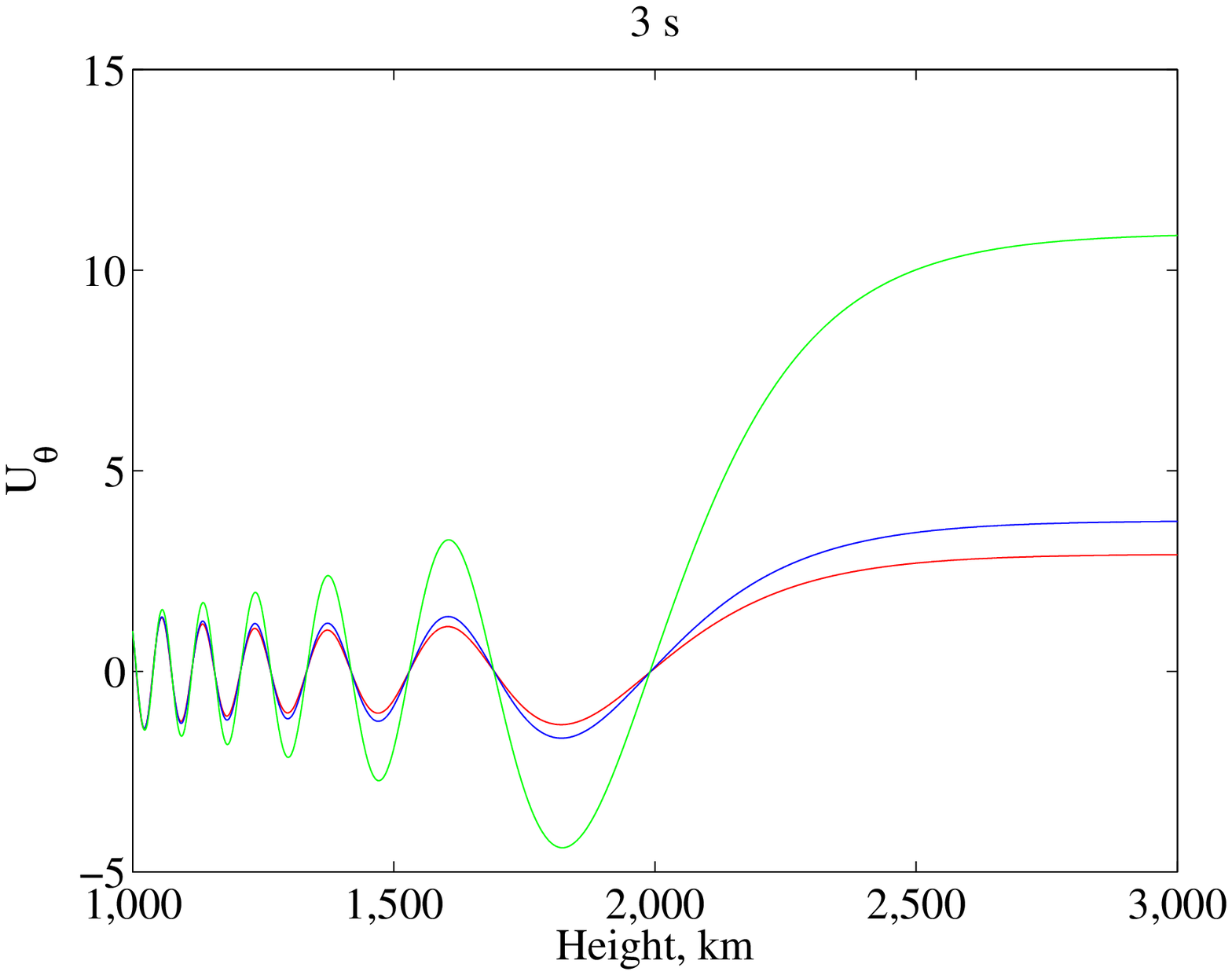}
\includegraphics[width=8.3cm]{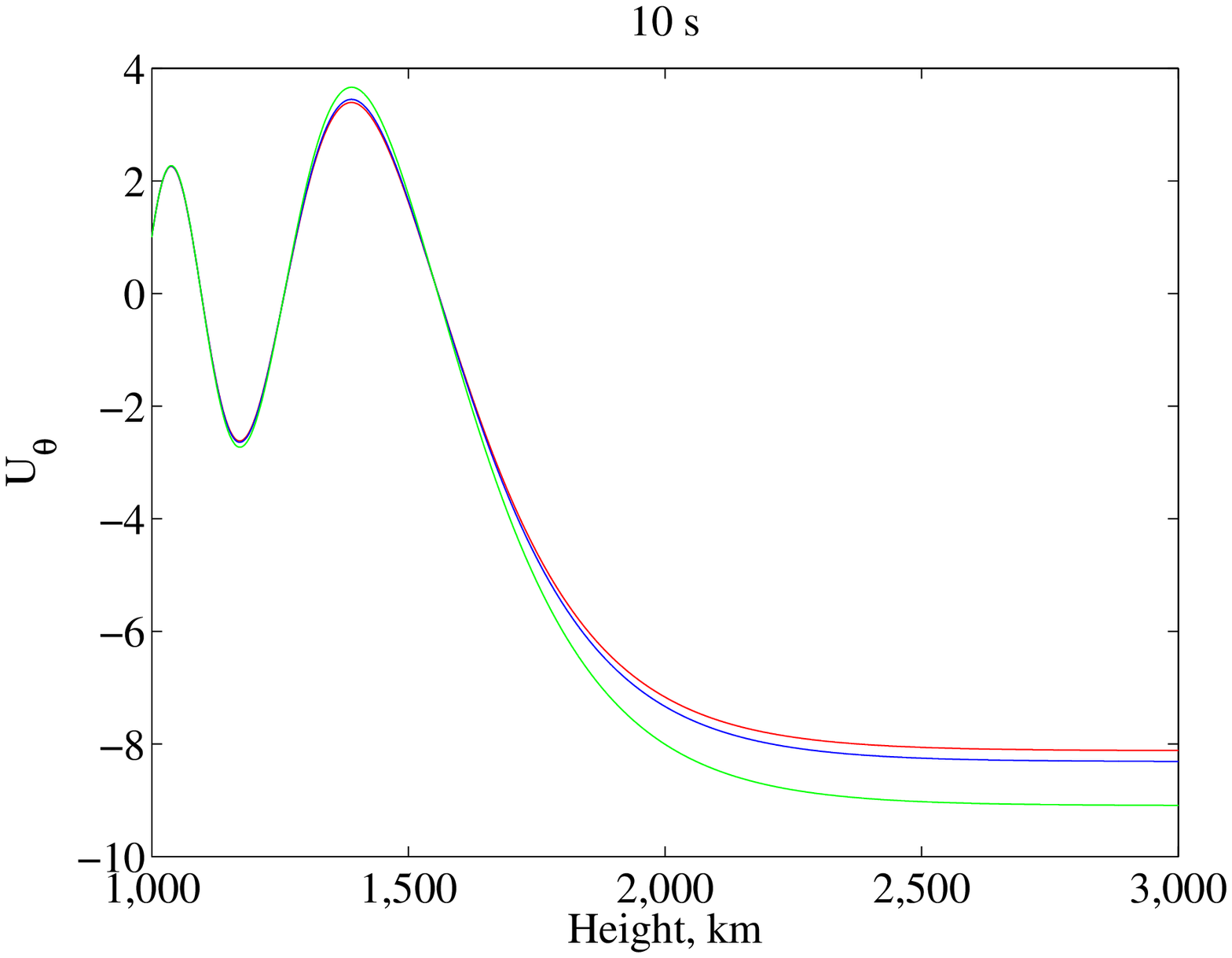}
\includegraphics[width=8.3cm]{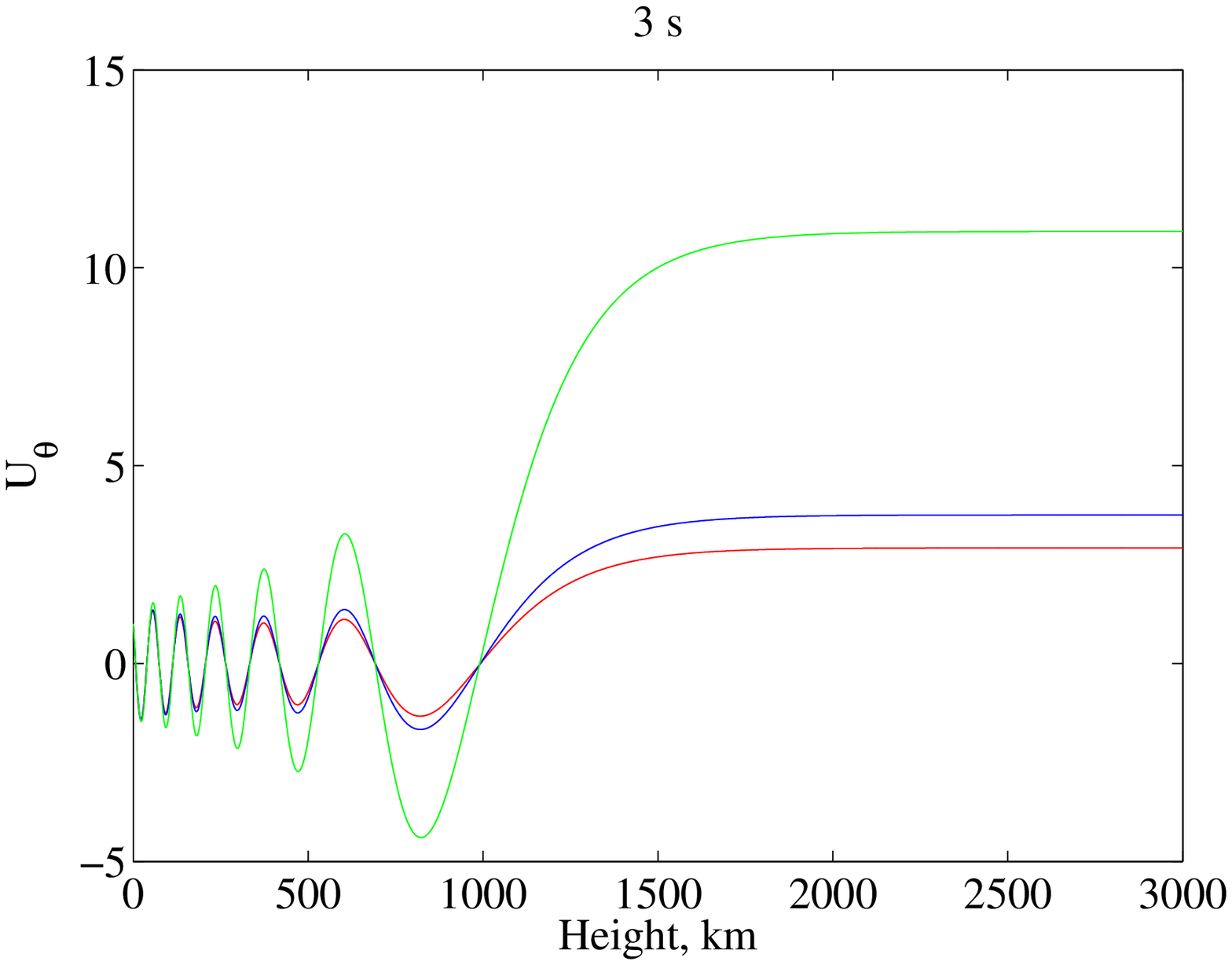}
\end{center}
\caption{Height dependence of steady state upward propagating torsional Alfv\'en waves with periods of 20 s (upper panel), 10 s (middle panel) and 3 s (lower panel). Green lines correspond to the fully ionized plasma, blue (red) lines correspond to partially ionized plasma with neutral hydrogen atoms (neutral hydrogen+neutral helium atoms).}
\end{figure}

Propagating waves are described by Hankel functions ($H^{1}_0(\zeta)=J_0(\zeta)+iY_0(\zeta)$, $H^{2}_0(\zeta)=J_0(\zeta)-iY_0(\zeta)$) and standing waves are described by Bessel functions ($J_0(\zeta)$). $H^{1}_0(\zeta)$ and $H^{2}_0(\zeta)$ functions have the same spatial dependence in the case of fully ionized plasma (Fig. 5). The upward and downward propagating waves can be distinguished by the sign of time-averaged Pointing flux (Hollweg \cite{Hollweg1984}), which shows that $H^{2}_0(\zeta)$ is upward propagating and $H^{1}_0(\zeta)$ is downward propagating waves. In the case of partially ionized plasma distinguishing between upward and downward propagating waves is complicated due to the complex argument of Hankel functions (De Pontieu et al. \cite{De Pontieu2001}). However, it is still possible to distinguish them by careful look into the height-structure of waves (Fig. 5). Red line, which corresponds to $H^{2}_0(\zeta)$, has smaller amplitude at higher heights comparing to the fully ionized case (green line), while blue line, which corresponds to $H^{1}_0(\zeta)$, has the stronger amplitude. This means that $H^{2}_0(\zeta)$ governs the wave, which damps at higher heights due to ion-neutral collisions, therefore it governs the dynamics of upward propagating waves. On the same basis, $H^{1}_0(\zeta)$ governs the downward propagating waves. The standing waves, expressed by $J_0(\zeta)$, can be formed after superposition of upward and downward propagating waves.

A mathematical proof that the function $H^{2}_0$ corresponds to upward waves is obtained from the asymptotic expansion of the Hankel functions for large arguments (see, e.g., Stenuit et al. \cite{Stenuit1999}). For large $\zeta$, which eventually means small $s$, the dependence of Hankel functions on $s$ can be written as
$$
H^{1}_0(\zeta) \sim \exp (- i |k_{s0}|s),\,\,H^{2}_0(\zeta) \sim \exp ( i |k_{s0}|s),
$$
where $k_{s0}$ is given in Equation~(\ref{ks0}). These expansions clearly show that $H^{2}_0$ corresponds to upward propagating waves when the temporal dependence is $\exp[-i \omega t]$. In the rest of the paper we consider only upward propagating waves. Note that only the real parts of $H^{1}_0(\zeta)$ and $H^{2}_0(\zeta)$, i.e., the part of the solution with the physical meaning, are shown in all  plots.

%
%\begin{rob2}
%The criterion based on Figure 5 that you use to choose which Hankel function represents upward waves looks  quite convincing but I wonder whether it is possible to give a mathematical proof. For example, using the asymptotic expansion of the Hankel functions for large arguments we should be able to write the Hankel functions in terms of complex exponentials. From there we could work out which function represents upward waves and which function represents downward waves. See, e.g., Stenuit et al. (1999, A\&A 342, 863).
%\end{rob2}
%
%\begin{rob}
% I imagine that in the Figures you are actually plotting the real part of the Hankel function. For this reason the result of $H^{1}_0(\zeta)$ and $H^{2}_0(\zeta)$ is the same for a fully ionized plasma because the real part of both functions is the same.  This should be mentioned.
%\end{rob}

Fig. 6 shows the height-dependence of upward propagating torsional Alfv\'en waves with different periods in the bright chromospheric network cores. At 1000 km height we use the number densities from FAL93-F model as $n_i$=3.22${\times}$10$^{11}$ cm$^{-3}$, $n_H$=2.65${\times}$10$^{13}$ cm$^{-3}$, $n_{He}$=2.68${\times}$10$^{12}$ cm$^{-3}$. The plasma temperature is set as $T$=8000 K and the Alfv\'en speed is set as $V_{A0}$= 20 km s$^{-1}$. It is seen that the behavior of longer period waves ( $>$ 10 s) is not significantly affected by ion-neutral collisions: the height dependence is similar for fully ionized (green line) and partially ionized (blue and red lines) plasmas (upper panel). The shorter period waves display more significant dependence on ion-neutral collision frequency. For example, the torsional Alfv\'en waves with period of 3 s are significantly damped in partially ionized plasma comparing to the fully ionized case (lower panel). It is also seen that the presence of neutral helium (red line) enhances the damping of Alfv\'en waves compared to neutral hydrogen only (blue line). Fig. 6 shows that the spatial dependence of waves are not oscillatory above particular heights i.e. waves become evanescent. In order to study this phenomenon, we consider wave propagation in fully ionized plasma.

Fig. 7 displays the velocity of torsional Alfv\'en waves vs height for different wave periods. It is seen that torsional Alfv\'en waves with period of 20 s become evanescent above 1500 km height. Waves with 5 s period become evanescent above 1800 km height, but the waves with 1 s period penetrate up to 2600 km height. Therefore, the longer period waves ( $>$ 5 s) do not reach to the transition region (2000 km height), but become evanescent at lower heights. This gravitational cut-off of Alfv\'en waves is known for a long time (Musielak and Moore, \cite{Musielak1995}). Recently, Murawski and Musielak (\cite{Murawski2010}) also showed that the linearly polarized Alfv\'en waves become evanescent above some heights in fully ionized plasma. Our result confirms their analyses.

\begin{figure}[t]
\vspace*{1mm}
\begin{center}
\includegraphics[width=8.5cm]{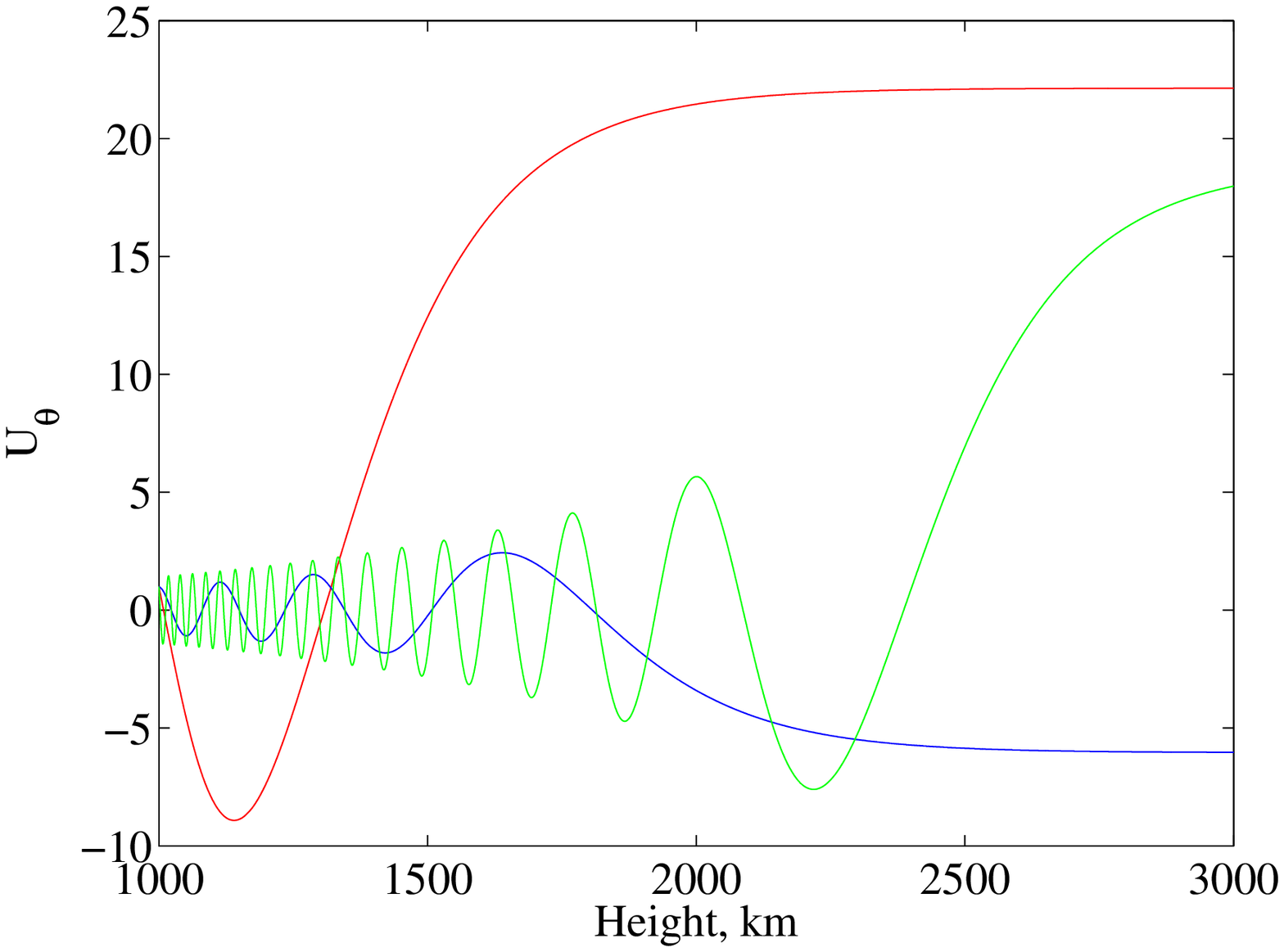}
\end{center}
\caption{Height dependence of steady state upward propagating torsional Alfv\'en waves with periods of 20 s (red line), 5 s (blue line) and 1 s
(green line) of different periods in fully ionized plasma.
}
\end{figure}

\section{Discussion}

Propagation of Alfv\'en waves in the chromosphere is very important for chromospheric/coronal heating as they may carry the photospheric energy into the upper layers. The magnetic field is concentrated in flux tubes at the photospheric level, therefore the torsional Alfv\'en waves can be generated due to vortex motions (Fedun et al. \cite{Fedun2011a}).  Observation of torsional Alfv\'en waves is possible through the periodic variation of spectral line width (Zaqarashvili \cite{Zaqarashvili2003}). Recently, Jess et al. (\cite{Jess2009}) reported the observation of torsional Alfv\'en waves in the lower chromosphere. The energy flux carried by the waves was enough to heat the solar corona. Therefore, it is interesting to study whether the waves may penetrate into the corona.

Here we study the torsional Alfv\'en waves in chromospheric magnetic flux tubes with partially ionized plasma taking into account neutral hydrogen and neutral helium atoms. We consider the stratification due to gravity, therefore the tubes are expanding with height. Hasan et al. (\cite{Hasan2003}) showed that the magnetic flux tubes can be considered as thin up to 1000 km height from the surface. Above this height, the neighboring magnetic flux tubes merge and the field lines are almost parallel. We use this model and split the magnetic tubes into two layers. We use the expanding thin flux tube approximation in the lower chromosphere below 1000 km. In the upper layer, above 1000 km we consider the magnetic field lines as vertical, so only medium density is changing with height. We use FAL93 model (Fontenla et al. \cite{Fontenla1993}) for the dependence of ion, neutral hydrogen and neutral helium number densities with height.

We start with three-fluid MHD description of partially ionized plasma, where one component is electron-proton-singly ionized helium and the two other components are neutral hydrogen and neutral helium gasses. Then we proceed to the single-fluid description by standard procedures and neglect the inertial terms of ion-neutral hydrogen and ion-neutral helium relative velocities. We obtain the Cowling diffusion coefficient (Eq. \ref{eqs3}), which is different than as previously used one. Namely, the expression Eq. (\ref{eqs3}) contains additional terms (the first and second terms in nominator) expressing the collision of ions with neutral hydrogen and neutral helium atoms separately. The previously used expression of Cowling diffusion coefficient is valid in weakly ionized plasma, where the coupling between neutral atoms are stronger.

In the lower part of magnetic flux tube, where the thin tube approximations is used, the dispersion relation for the torsional Alfv\'en waves is obtained. The solution of the dispersion relation shows that the damping of Alfv\'en waves due to ion-neutral collision is significant in faint cell center areas at all frequencies, while only high-frequency waves are damped in the chromospheric network cores (Fig. 4). This means that torsional Alfv\'en waves may propagate without problem up to 1000 km above the surface in chromospheric network cores. The effect of neutral helium atoms is important in all cases and it increases for lower frequency waves.

In the upper part of magnetic flux tube, the height dependence of torsional Alfv\'en wave velocity is governed by Bessel-type equation, therefore
the solutions are Bessel or Hankel functions with zero order and complex arguments. We consider only upward propagating waves, which are expressed by Hankel functions, $H^{2}_0$. The solution of steady-state torsional Alfv\'en waves shows that the long-period waves are not affected by ion-neutral collisions. Only short-period waves ($<$ 5 s) are damped significantly (Fig. 6). Existence of neutral helium atoms enhances damping of the waves but not significantly. On the other hand, long-period waves show no oscillatory behavior above certain heights, so they become evanescent in upper part of the chromosphere. This phenomenon is called gravitational cut-off (Musielak et al. \cite{Musielak1995}, Murawski \& Musielak \cite{Murawski2010}). It is seen that only the waves with very short period ($\sim$ 1 s) may reach the transition region in the case of fully ionized plasma (Fig. 7). This means that the low-frequency waves can not reach the transition region due to the gravitational cut-off, but high-frequency waves are damped due to ion-neutral collision. Then the torsional Alfv\'en waves can not reach the transition region at all. This is true near the axis of magnetic flux tubes, where our approximations are valid. However, it is possible that some wave tunneling exists there and the evanescent tail of the waves can indeed reach the transition region and the corona. Due to the change of the ambient conditions the waves may become propagating again in the corona. In the outer part of tubes, where magnetic field lines are more inclined, the problem of wave propagation along the chromosphere may not arise at all. It requires further detailed study.

On the other hand, Alfv\'en waves may easily penetrate into the corona if they are excited in the higher part of the chromosphere (Hansen and Cally \cite{Hansen2012}). Long-period acoustic oscillations (p-modes) are evanescent in the lower atmosphere due to the gravitational stratification, however they may propagate with an angle about the vertical and may trigger Alfv\'en waves in the upper chromosphere through mode conversion (Zaqarashvili and Roberts \cite{Zaqarashvili2006}, Cally and Goossens \cite{Cally2008}, Cally and Hansen \cite{Cally2011}, Khomenko and Cally \cite{Khomenko2012}). If low-frequency torsional Alfv\'en waves are excited above 1000 km height, then our model also allows the propagation of the waves into the corona as the evanescent point can be shifted up above the transition region (Fig. 7).

Recently, Vranjes et al. (\cite{Vranjes2008}) suggested that the energy flux of Alfv\'en waves excited in the photosphere is overestimated and thus the Alfv\'en waves can be hardly excited in the photosphere (but see Tsap et al. \cite{Tsap2011} for alternative point of view). This is indeed an interesting question and should be addressed adequately. One possible explanation is that Vranjes et al. (\cite{Vranjes2008}) considered the magnetic field strength of 100 G in the photosphere, which is too weak for flux tubes, where the magnetic field strength is of order kG. Therefore, it is possible the Alfv\'en waves are excited inside flux tubes only, but not outside the tubes, where the magnetic field is relatively weak. This point needs further discussion.

Ion-neutral collisions are important only for short-period waves ($<$ 5 s). On the other hand, only long period transverse oscillations ($\geq$ 3 min) are frequently observed so far in the solar atmosphere. However, some observations from ground based coronagraphs show the oscillations of spicule axis with periods of $\leq$ 1 min (Zaqarashvili et al. \cite{Zaqarashvili20071}, Zaqarashvili \& Erd{\'e}lyi \cite{Zaqarashvili2009}). Several reasons could be responsible for the absence of high-frequency oscillations: not enough tempo-spatial resolution of observations, hard excitation in the photospheric level, quick damping in the lower atmosphere etc. We hope that more sophisticated observations will help us to improve our knowledge about high-frequency oscillations in the future.

%
%\begin{rob}
%It is possible that there is some wave tunnelling and the evanescent tail of the waves can indeed reach the transition region and the corona. Due to the change of the ambient conditions the waves may become propagating again in the corona.
%\end{rob}

\section{Conclusions}

Propagation of torsional Alfv\'en waves along expanding vertical magnetic flux tubes in the partially ionized solar chromosphere is studied taking into account the ion collisions with neutral hydrogen and neutral helium atoms. The waves propagate freely in the lower chromosphere up to 1000 km and may damp by ion-neutral collisions. The new expression of the Cowling diffusion (and consequently damping rate) including the neutral helium is obtained (Eq. 15). Neutral helium atoms may significantly enhance the damping of Alfv\'en waves. On the other hand, the long period ($>$ 5 s) waves propagating near the tube axis become evanescent above some height in the upper chromosphere not reaching the transition region, while the short period waves ($<$ 5 s) are damped by ion-neutral collisions. Hence the torsional Alfv\'en waves may not reach the transition region and consequently the solar corona unless some tunneling effects are considered. This means that the coronal heating by photospheric Alfv\'en waves should be considered by caution. At the same time, energy of the waves can be dissipated in the chromosphere leading to the heating of ambient plasma.

\begin{acknowledgements}
The work was supported by the Austrian Fonds zur F\"orderung
der wissenschaftlichen Forschung (project P21197-N16) and by the European FP7-PEOPLE-2010-IRSES-269299 project- SOLSPANET.
RS acknowledges support from a Marie Curie Intra-European Fellowship within the European Commission 7th Framework Program  (PIEF-GA-2010-274716). RS also acknowledges support from MINECO and FEDER funds through project AYA2011-22846 and from CAIB through the `Grups Competitius' scheme.
\end{acknowledgements}

\appendix

\end{document}